\documentclass[pra,twocolumn,showpacs,nofootinbib,floatfix]{revtex4}

\usepackage{amsmath,amsfonts,amssymb,bm}
\usepackage{dcolumn}
\usepackage[final]{graphicx}
\usepackage{bm,tabularx}

\begin{document}

\title{Optimal quantum control of Bose-Einstein condensates in magnetic microtraps: Consideration of filter effects}

\author{Georg J\"ager}
\author{Ulrich Hohenester}
\affiliation{Institut f\"ur Physik, Karl-Franzens-Universit\"at Graz, Universit\"atsplatz 5, 8010 Graz, Austria}

\date{September 7, 2013}

\begin{abstract}
We theoretically investigate protocols based on optimal control theory (OCT) for manipulating Bose-Einstein condensates in magnetic microtraps, using the framework of the Gross-Pitaevskii equation.  In our approach we explicitly account for filter functions that distort the computed optimal control, a situation inherent to many experimental OCT implementations.  We apply our scheme to the shakeup process of a condensate from the ground to the first excited state, following a recent experimental and theoretical study, and demonstrate that the fidelity of OCT protocols is not significantly deteriorated by typical filters.
\end{abstract}


\maketitle

Optimal quantum control aims at the manipulation of a quantum-mechanical wavefunction in a controlled fashion~\cite{peirce:88,rabitz:00,brif:10}.  External parameters, such as laser fields, can be controlled at will, and allow to steer the wavefunction from a given initial to a desired terminal state.  Recent years have seen tremendous research efforts in the realm of quantum control~\cite{brif:10}.  Quantum chemistry implementations often rely on stochastic optimization techniques, which are particulary appealing for experimental implementations~\cite{rabitz:00}.  An alternative approach is provided by optimal control theory (OCT)~\cite{tannor:85,peirce:88,zhu:98}, that performs a numerical optimization of the control fields through an iterative procedure by solving the dynamic system equations.

In Ref.~\cite{buecker:11,buecker:13} we have presented an experimental implementation of optimal quantum control for a Bose-Einstein condensate.  Ultracold atoms become trapped in the vicinity of an atom chip~\cite{reichel:11} by the magnetic fields produced by currents running through the wires of the chip, and the magnetic confinement potential can be controlled by changing the currents.  We have demonstrated the excitation of the condensate wavefunction from the ground to the first excited state of an anharmonic potential, where the population transfer has been achieved with an efficiency close to 100\% by displacing the potential minimum according to a protocol computed with optimal control theory.

In this Brief Report, we investigate the effects of filter functions that distort the control parameters computed from optimal control theory.  Such filters might be due to electronics and are inherent to many experiments.  For sufficiently simple control protocols filter effects can be corrected through a simple deconvolution scheme, but in general it is advantageous to incorporate filtering directly in the OCT approach.  In this paper we first develop the methodology for OCT with filtered control parameters, and then apply our scheme to the condensate shakeup investigated in Refs.~\cite{buecker:11,buecker:13}.  We find that for realistic filter functions the fidelity of the control process does not become deteriorated significantly.  Although in this paper we only focus on Bose-Einstein condensates, the developed methodology is general and might be useful in a much wider context.

\textit{OCT without filter}.---We first briefly review the optimal control implementation of Bose-Einstein condensates formulated in Ref.~\cite{hohenester.pra:07} (see also Ref.~\cite{sklarz:02} for related work).  Within the framework of the Gross-Pitaevskii equation~\cite{dalfovo:99,leggett:01} the dynamics of the condensate wavefunction $\psi(\bm r,t)$ is described by ($\hbar=1$)
\begin{equation}\label{eq:gp}
 i\hbar\frac{\partial\psi(\bm r,t)}{\partial t}=
   \Bigl(-\frac{\nabla^2}{2M} + V(\bm r,\lambda(t)) + 
   \kappa\bigl|\psi(\bm r,t)\bigr|^2\Bigr)\psi(\bm r,t)\,.
\end{equation}
The first term on the right-hand side is the operator for the kinetic energy, the second one is the confinement potential that can be controlled by some external parameter $\lambda(t)$, and the last term is the nonlinear atom-atom interaction in the mean field approximation of the Gross-Pitaevskii framework.  $M$ is the atom mass and $\kappa$ is the strength of the atom-atom interactions. 

Optimal control theory is seeking for an ``optimal'' time variation of the control parameter $\lambda(t)$ in order to fulfill certain control objectives.  For instance, the cost function
\begin{equation}\label{eq:cost}
  J(\psi,\lambda)= \frac{1}{2} \left[1-\left| \left< \psi_d |\psi(T)\right> \right |^2 \right]  +
    \frac{\gamma}{2} \int_0^T[\dot{\lambda}(t)]^2 dt\,
\end{equation}
becomes minimal when the state $\psi(T)$ at the terminal time $T$ of the control process comes as close as possible to a \textit{desired} state $\psi_d$, apart from an irrelevant global phase \cite{hohenester.pra:07}.  The second term penalizes strong variations of the control parameter and is needed to make the OCT problem well posed~\cite{hohenester.pra:07,borzi:08,vonwinckel:08}.  Through $\gamma$ it is possible to weight the importance of wavefunction matching and control smoothness, and below we will set $\gamma\ll 1$.  

\begin{figure*}
\centerline{\includegraphics[width=1.5\columnwidth]{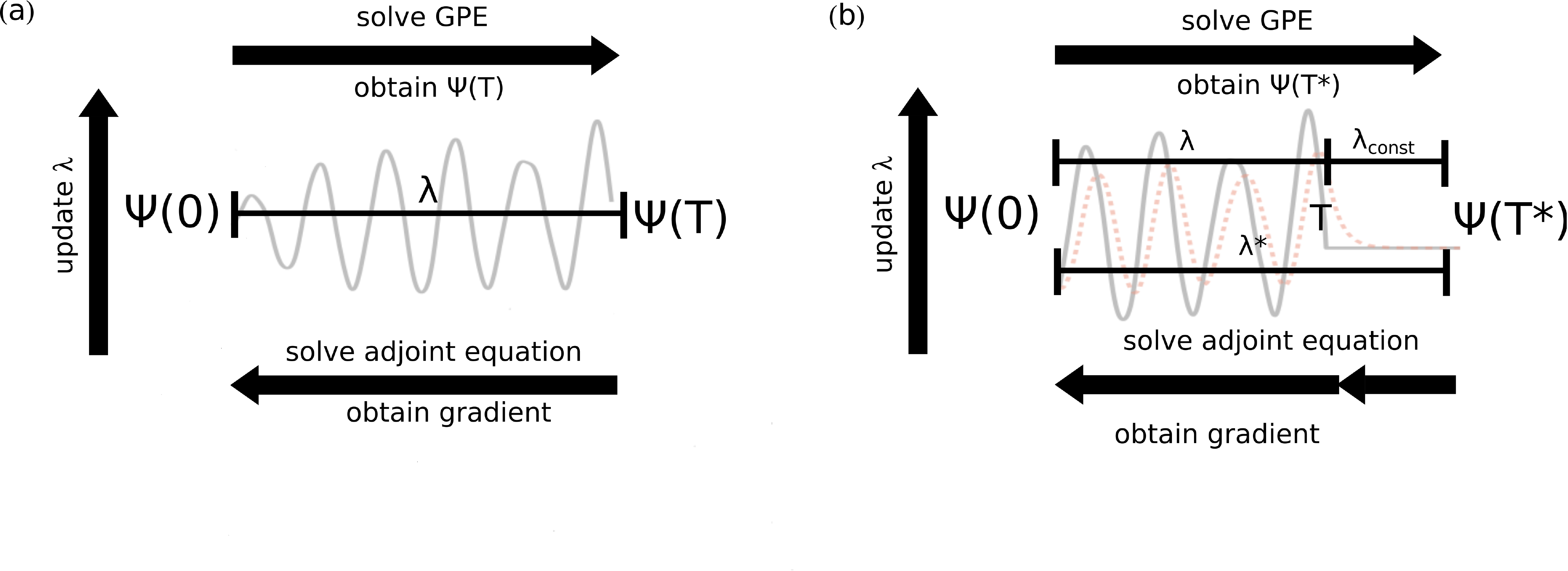}}
\vspace*{-0.5cm}
\caption{(Color online) Optimization strategy (a) without and (b) with a filter function.  When a filter is considered, the optimization interval $[0,T]$ should differ from the time interval $[0,T^\star]$ of the condensate dynamics to account for the finite filter response time.}
\end{figure*}

In order to bring the system from the initial state $\psi_0$ to the terminal state $\psi(T)$ we have to fulfill the Gross-Pitaevskii equation, which enters as a \textit{constraint} in our optimization problem.  The constrained optimization problem can be turned into an unconstrained one by means of Lagrange multipliers $p(t)$.  To this end, we introduce a Lagrange function
\begin{eqnarray}\label{eq:lagrange}
  &&L(\psi,p,\lambda)=J(\psi,\lambda)\\ &&\quad+\mbox{Re}\Biggl[
  \int_0^T \Bigl< p(t)\Bigr|i\dot\psi(t)-
  \left(H_\lambda+\kappa|\psi(t)|^2\right)\psi(t)\Bigr> dt\Biggr]\,,\nonumber
\end{eqnarray}
where $H_\lambda$ is the single-particle Hamiltonian defined through Eq.~\eqref{eq:gp}.  The Lagrange function has a saddle point at the minimum of $J(\psi,\lambda)$ where all derivatives $\delta L/\delta\psi^*$, $\delta L/\delta p^*$, and $\delta L/\delta\lambda$ become zero.  Performing
functional derivatives in the Lagrange function, we then arrive at the following set of equations~\cite{hohenester.pra:07,borzi:08}
\begin{subequations}\label{eq:optimality}
\begin{eqnarray}
   i\dot\psi&=&\left(-\frac{\nabla^2}{2M} + V(\bm r,\lambda(t))+\kappa|\psi|^2\right)\psi
  \label{eq:opt1}\\
   i\dot p&=&\left(-\frac{\nabla^2}{2M} + V(\bm r,\lambda(t))+2\kappa|\psi|^2\right)p+
   \kappa\psi^2 p^* \label{eq:opt2}\,\,\\
   \gamma\ddot\lambda&=&-\mbox{Re}\Bigl<p\Bigr|
   \frac{\partial H_\lambda}{\partial\lambda}\Bigl|\psi\Bigr>\,.\label{eq:opt3}
\end{eqnarray}
\end{subequations}
Eq.~\eqref{eq:opt1} is the initial value problem $\psi(0)=\psi_0$ of the Gross-Pitaevskii equation, whereas Eq.~\eqref{eq:opt2} is a terminal value problem for the adjoint variable $p(T)=i\langle\psi_d|\psi(T)\rangle\psi_d$.  Finally, Eq.~\eqref{eq:opt3}  determines the optimal control and is a boundary value problem where both the initial and terminal value are fixed, $\lambda(0)=\lambda_0$ and $\lambda(T)=\lambda_T$.  It has been discussed in Refs.~\cite{hohenester.pra:07,vonwinckel:08,grond.pra:09b} that Eqs.~(\ref{eq:optimality}a--c) can be also used for arbitrary control parameters in order to formulate an iterative procedure that successively improves $\lambda(t)$.

\begin{figure*}
\centerline{\includegraphics[width=1.72\columnwidth]{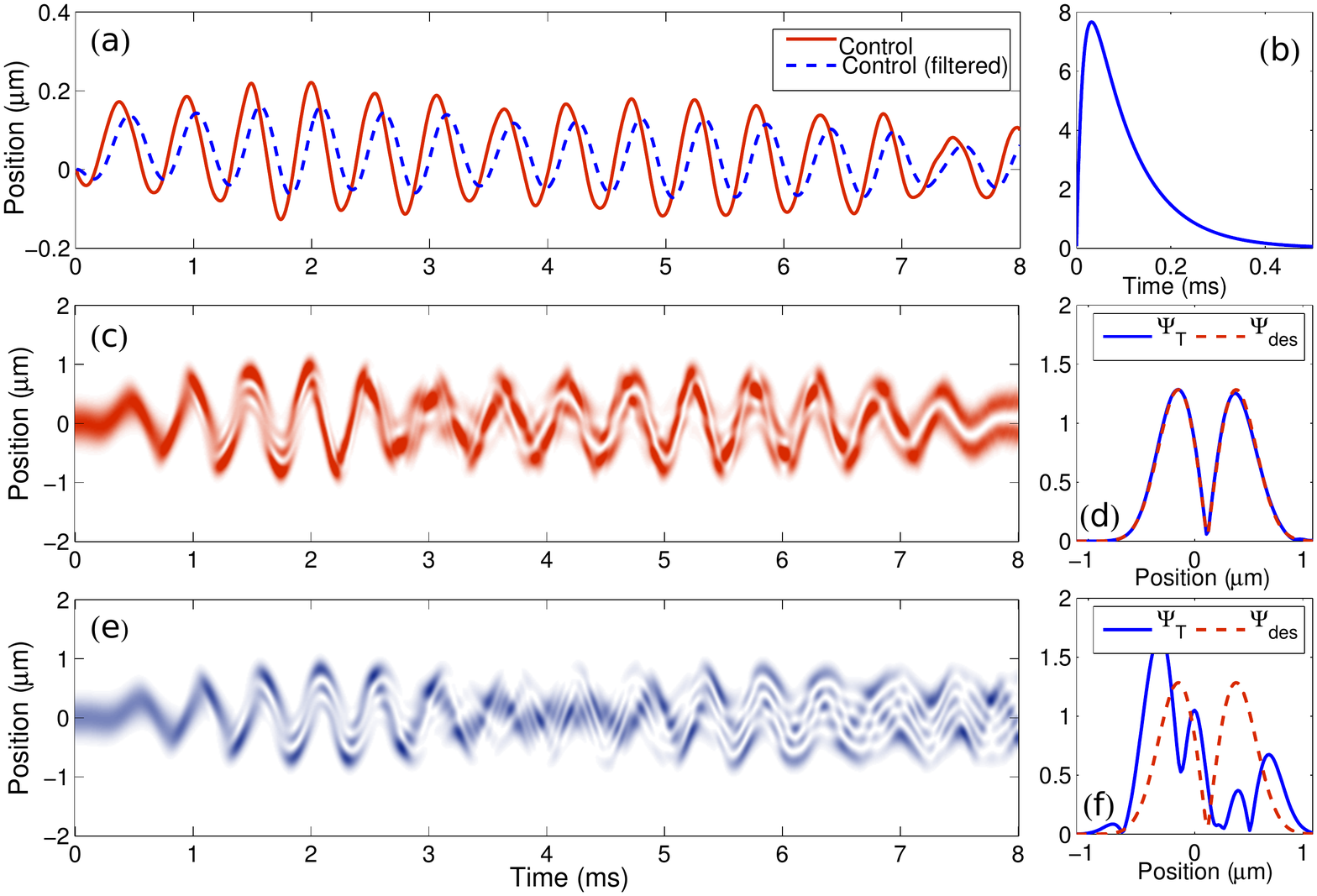}}
\vspace*{-0.5cm}
\caption{(Color online) (a) Filtered and unfiltered control for $\lambda$ obtained \textit{without} the consideration of a filter. (b) Typical filter function $h(t)$.  (c) Time evolution of the condensate density under the effect of the unfiltered control.  (d) Absolute value of desired and terminal wave function for unfiltered control at terminal time $T$ of control process. (e,f) Same as panels (c,d) but for the filtered control $\lambda^\star(t)$.}
\centerline{\includegraphics[width=1.72\columnwidth]{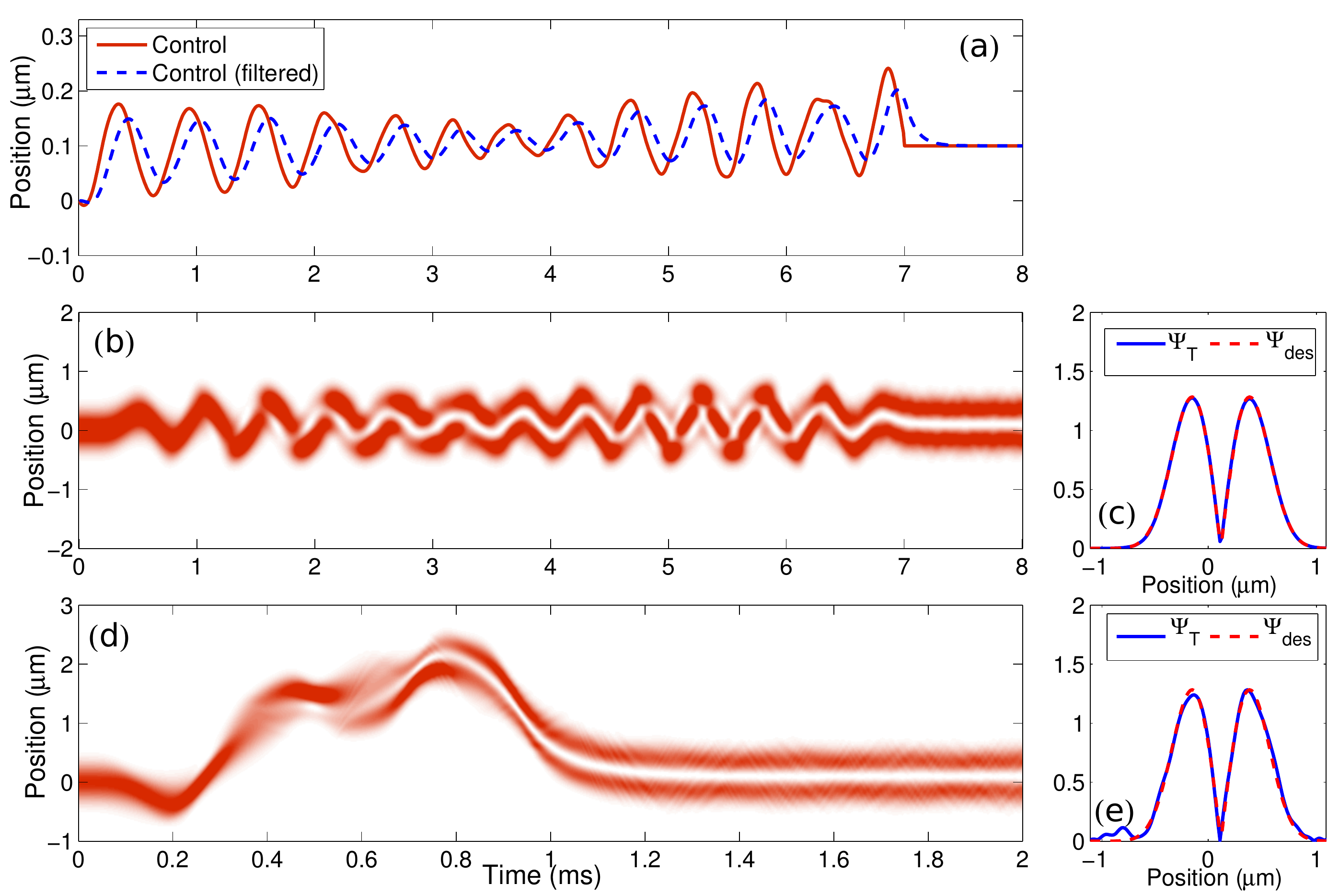}}
\caption{(Color online) (a) Optimal control \textit{with} consideration of the a filter function. The last part of the time evolution of $\lambda$ is kept constant in order to guarantee $\lambda^\star(T^\star)=\lambda_T$. (b) Time development of the condensate density driven by the optimal control for a transition time of 8 ms. (c) Absolute value of desired and terminal wave function at final time.  (d,e) Same as panels (b,c), but for a shorter transition time of 2 ms.}
\end{figure*}

\textit{OCT with filter}.---In this paper we discuss the case that the external control parameter $\lambda(t)$ does not directly influence the confinement potential, but becomes distorted by some filter function $h(\tau)$.  Such filtering is inherent to many quantum control experiments, and has previously been also discussed in the context of laser pulse shaping~\cite{sundermann:00,gollub:08}.  Fig.~2(b) shows a typical filter function due to the finite response time of the electronics.  As a consequence, when the optimized control $\lambda(t)$ is sent to the electronics, a distorted (non-optimal) signal
\begin{equation}\label{eq:conv}
  \lambda^{\star}(t) = \int_0^t h(\tau) \lambda(t - \tau)\,d \tau
\end{equation}
determines the time evolution of $V(\bm r,\lambda^\star(t))$.  It is obvious that the condensate dynamics with the non-optimized $\lambda^\star(t)$ will no longer bring the condensate wavefunction to the desired state.

A possible solution is to try to deconvolute Eq.~\eqref{eq:conv} and to find an input signal that produces the proper output signal.  However, since $h$ is typically a filter for high frequencies, and the optimal control might include fast time variations, the deconvolution is often doomed to failure.  In what follows we formulate a different strategy.  We stay with the OCT framework of Eqs.~(\ref{eq:gp}--\ref{eq:lagrange}), but replace $\lambda(t)$ with the filtered $\lambda^\star(t)$.  Our task now is to determine the optimal control $\lambda(t)$ such that the filtered control of Eq.~\eqref{eq:conv} brings the condensate from $\psi_0$ to $\psi_d$.  In fact, we can carry over most of the results of the previous discussion.  However, in Eqs.~(\ref{eq:opt1},\ref{eq:opt2}) we have to replace $\lambda(t)$ with $\lambda^\star(t)$, and the control Eq.~\eqref{eq:opt3} is changed to a form
\begin{equation}\label{eq:opt3star}
  \gamma\ddot{\lambda} = -\mbox{Re} \int_t^T h(s-t) \left\langle  p(s)\left|  \frac{\partial V}{\partial\lambda}\right|\psi(s) \right\rangle\, ds
\end{equation}
that is now non-local in time (note that $\partial V/\partial\lambda$ is evaluated at time $s$).  The optimality system is then formed by Eqs.~(\ref{eq:optimality}a,b), with $\lambda$ replaced by $\lambda^\star$, together with Eq.~\eqref{eq:opt3star}.

A slight complication appears at this point.  Let us consider the schematic OCT loop depicted in Fig.~1.  In the unfiltered case of panel (a) the optimal control parameter $\lambda(t)$ determines how the condensate wavefunction is brought to $\psi_d$.  If $\psi_d$ is a stationary state of the Gross-Pitaevskii equation and $\lambda$ is kept fixed at $t\ge T$, the system will remain in this stationary state.  On the other hand, in presence of a filter things behave differently.  For $\lambda(t\ge T)=\lambda_T$ the filtered response $\lambda^\star(t)$ can still vary at times later than $T$, because of the finite response time of the filter.  Thus, even if the system ends up in the desired state $\psi(T)=\psi_d$ at the terminal time $T$, the ensuing temporal evolution of $\lambda^\star$ will push the system away from $\psi_d$.

To account for this, we propose a slight variation of our OCT implementation.  We use different time intervals $[0,T]$ and $[0,T^\star]$ for the control optimization and the condensate simulation, respectively.  The two end points differ by $T^\star-T=\tau^\star$, as shown in Fig.~1(b).  $\tau^\star$ is a time where the filter function has dropped to zero (approximately the inverse of the cutoff frequency).  Thus, if $\lambda$ is kept constant at times $t\ge T$, the filterd response $\lambda^\star$ will become constant at $t\ge T^\star$.  

In presence of filtering, the OCT loop formulated in Refs.~\cite{hohenester.pra:07,borzi:08,vonwinckel:08} consists of the following steps:

\medskip
\begin{tabularx}{0.95\columnwidth}{cX}
1. & Start with some initial guess for $\lambda(t)$, where $\lambda$ is kept constant for $t\ge T$. \\
2. & Solve the Gross-Pitaevskii Eq.~\eqref{eq:opt1} forwards in the time interval $[0,T^\star]$ using the filtered control $\lambda^\star$. \\
3. & Determine the terminal condition $p(T^\star)$ for the adjoint variable, and integrate the sensitivity Eq.~\eqref{eq:opt2} backwards in time.\\
4. & Compute a search direction for $\lambda(t)$ via Eq.~\eqref{eq:opt3star}. \\
5. & Compute an improved control parameter $\lambda(t)$ within the time interval $[0,T]$, and iterate the loop 2--5 until convergence is obtained or a given number of iterations is reached.
\end{tabularx}

\textit{Results}.---We next apply the OCT implementation including filter functions to the case of condensate shakeup described in Refs.~\cite{buecker:11,buecker:13}.  Here one starts in the groundstate of an anharmonic trap, and the trap center is displaced in an optimized fashion such that the condensate is brought to the first excited state.  We only consider the spatial dynamics along the diplacement direction, and use the same trap and condensate parameters as in Refs.~\cite{buecker:11,buecker:13}.  The solid line in Fig.~2(a) shows the optimized control for a shakeup process, using a time interval of 8 ms and without filter effects.  Here $\lambda(t)$ directly corresponds to the displacement of the trap center.  The density plot in panel (c) provides details about how the condensate is brought from the initial to the desired state.  Panel (d) demonstrates that $\psi(T)$ and $\psi_d$ almost perfectly match at the terminal time $T$.  In presence of a filter function, depicted in panel (b), the control parameter becomes distorted [dashed line in panel (a)] and the control process leaves the system in a highly excited and non-stationary state, as shown in Figs.~2(e,f).

\begin{figure}
\centerline{\includegraphics[height=0.8\columnwidth,angle=90]{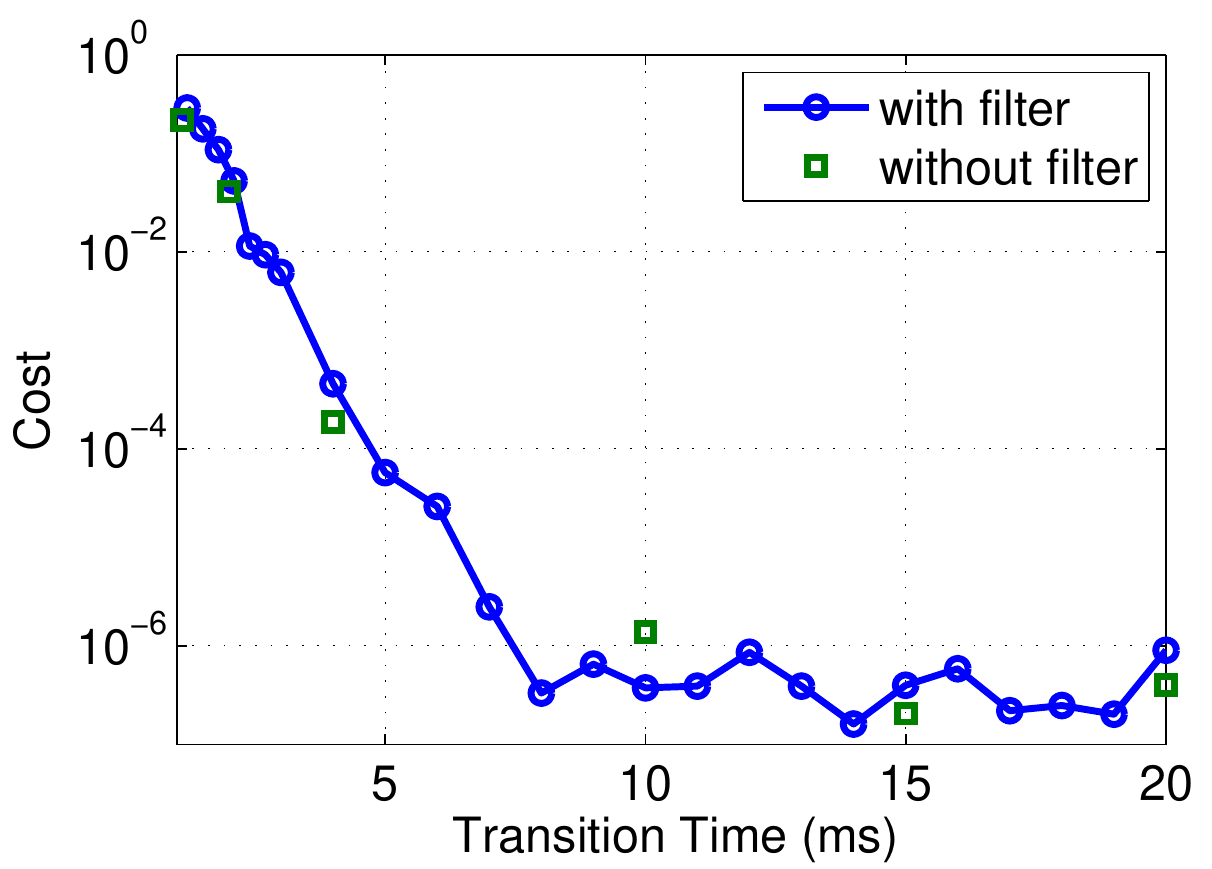}} 
\caption{(Color online) Final cost for different transition times (we set $\gamma=10^{-9}$). For transition times under 8 ms the final cost begins to rise, for transition times under 2 ms no satisfying control could be obtained.}
\end{figure}

Things change considerably for the OCT implementation with filtering, introduced above.  Fig.~3(a) shows the optimized control $\lambda(t)$ together with the filtered signal $\lambda^\star(t)$.  In our OCT simulations we set the filter response time $\tau^\star$ to 0.5 ms.  Panels (b,c) show that with the optimized control parameter the desired state is reached perfectly, even in presence of filtering.  In panel (d) we finally demonstrate that the shakeup protocol also works for significantly shorter transition times, here 2 ms.  In Fig.~4 we investigate the success for the shakeup process for different times $T^\star$.  For small transition times, say around 2 ms, the transfer has a high efficiency but the terminal state still somewhat differs from the desired one.  With increasing transition time the efficiency increases, and the cost function saturates at later times.

In conclusion, we have developed a methodology that allows to incorporate filter effects in optimal quantum control simulations.  We have applied our approach to a condensate shakeup process, where the wavefunction is transferred from the ground to the first excited state of an anharmonic, magnetic microtrap, and have demonstrated that high transfer efficiencies can be achieved even in presence of filtering.  Although in this paper we have only focused on Bose-Einstein condensates, the developed methodology is rather general and might be also useful for other systems.

\textit{Acknowledgments}.---We are grateful to J\"org Schmiedmayer, Robert B\"ucker, Tarik Berrada, and Thorsten Schumm for most helpful discussions and for providing us with the electronic filter data.  This work has been supported in part by the Austrian science fund FWF under project P24248.

\end{document}